\documentclass[final,3p,times]{elsarticle}

\usepackage{filecontents}

\usepackage{tikz}
\def\checkmark{\tikz\fill[scale=0.4](0,.35) -- (.25,0) -- (1,.7) -- (.25,.15) -- cycle;}
\usepackage{makecell}
\usepackage{hhline}
\usepackage{amsmath}
\usepackage{xurl}
\usepackage{enumitem}
\usepackage{soul}

\usepackage[normalem]{ulem}

\usepackage [english]{babel}
\usepackage [autostyle, english = american]{csquotes}
\MakeOuterQuote{"}

\usepackage{gensymb}

\usepackage{framed} 

\usepackage{multicol} 

\usepackage{nomencl} 

\makenomenclature

\renewcommand*\nompreamble{\begin{multicols}{2}}

\renewcommand*\nompostamble{\end{multicols}}

\usepackage{etoolbox}

\usepackage{amssymb}
 \usepackage{amsthm}
\usepackage{cuted}

\journal{Elsevier}

\begin{document}

\begin{frontmatter}



\title{Levelised Cost of Demand Response: Estimating the Cost-Competitiveness of Flexible Demand}


\iftrue
\author[inst1,inst2]{Jacob Thrän\corref{cor1}}
\ead{j.thran22@imperial.ac.uk}
\cortext[cor1]{Corresponding author}

\affiliation[inst1]{organization={Department of Electrical and Electronic Engineering, Imperial College London},
            addressline={South Kensington Campus}, 
            city={London},
            postcode={SW7 2AZ}, 
            country={United Kingdom}}

\author[inst1]{Tim C. Green}
\author[inst2]{Robert Shorten}

\affiliation[inst2]{organization={Dyson School of Design Engineering, Imperial College London},
            addressline={South Kensington Campus}, 
            city={London},
            postcode={SW7 2AZ}, 
            country={United Kingdom}}
\fi
\begin{abstract}
To make well-informed investment decisions, energy system stakeholders require reliable cost frameworks for demand response and storage technologies. While the levelised cost of storage permits comprehensive cost comparisons between different storage technologies, no generic cost measure for the comparison of different demand response schemes exists. This paper introduces the levelised cost of demand response, which is an analogous measure to the levelised cost of storage but crucially differs from it by considering consumer reward payments. Additionally, the value factor from cost estimations of variable renewable energy is adapted to account for the variable availability of demand response. The levelised cost of demand response is estimated for four direct load control schemes and twelve storage applications, and then contrasted against literature values for the levelised cost of the most competitive storage technologies. The direct load control schemes are vehicle-to-grid, smart charging, smart heat pumps, and heat pumps with thermal storage. The results show that only heat pumps with thermal storage consistently outcompete storage technologies, with EV-based schemes being competitive for some applications. The results and the underlying methodology offer a tool for energy system stakeholders to assess the competitiveness of demand response schemes even with limited user data.
\end{abstract}

\iftrue
\begin{graphicalabstract}
\includegraphics[width=1.0\textwidth]{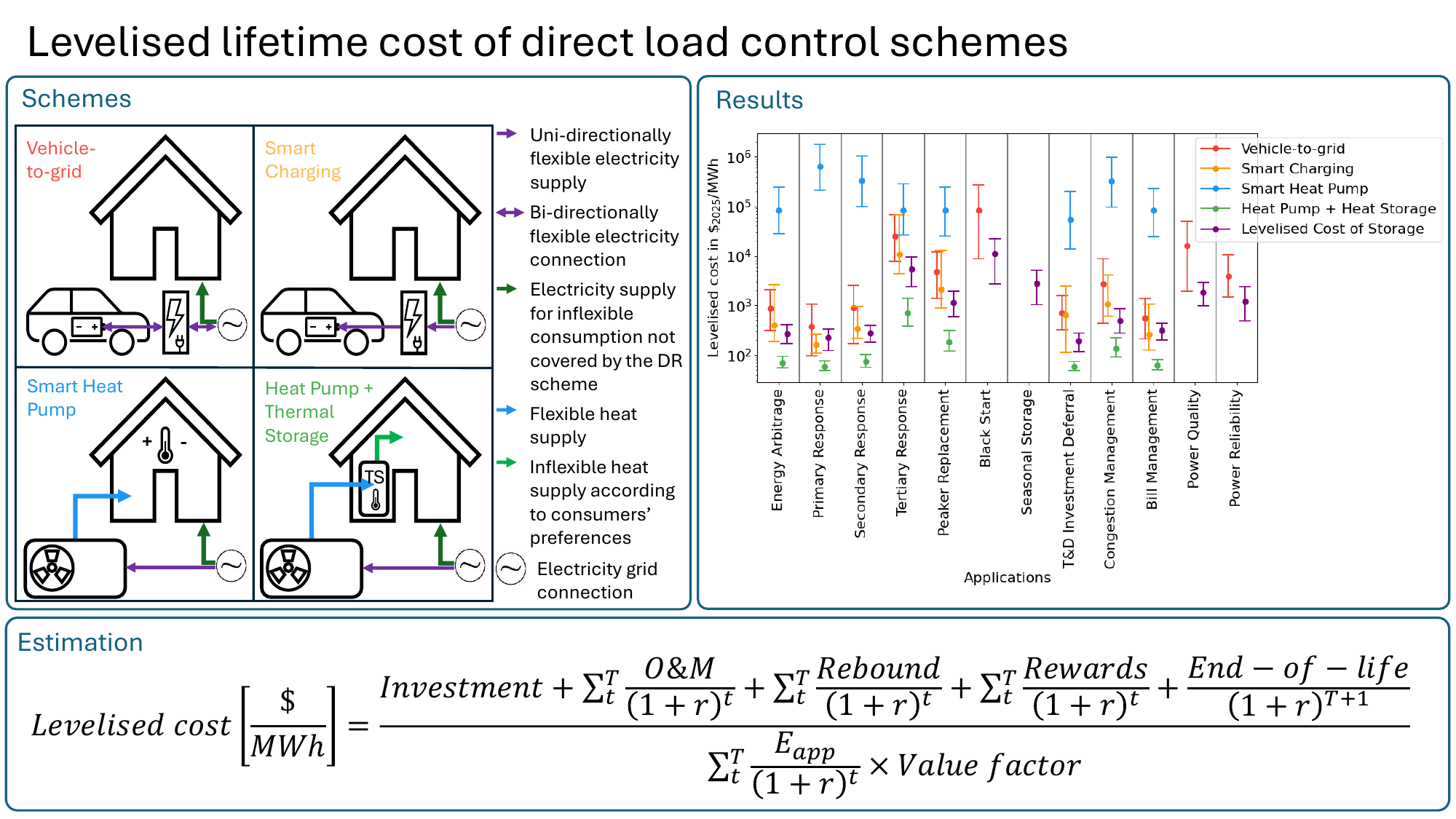} 
\end{graphicalabstract}
\fi
\begin{highlights}
\item The levelised cost of demand response compares the costs of load control and storage
\item The framework includes consumer payments based on stated-choice experiment literature
\item Schemes: vehicle-to-grid, smart charging, smart heat pump, heat pump + heat storage
\item Heat pump + heat storage is cheaper than any other demand response or storage option
\item Vehicle-to-grid and smart charging are competitive with storage for some applications
\end{highlights}


\end{frontmatter}


\section{Introduction}
\label{sec:introduction}


Demand response (DR) has long been praised for its potential benefits for power system operation \citep{strbac2008demand}. The energy transition, with its need for the integration of increasing shares of variable renewable energy (VRE), has only amplified these potential benefits \citep{gils2016economic}. The addition of large domestic loads, like electric vehicles (EVs) and heat pumps (HPs), has simultaneously increased the effectiveness of DR \citep{fischer2017heat}, and advances in digital control have reduced the perceived inconvenience associated with it \citep{faruqui2013arcturus}. Barriers to DR remain, though, with two regularly mentioned challenges being the fragmented business case of DR and the regulatory framework around it \citep{pinson2014benefits}. The business case for DR is described as fragmented because it has the potential to create benefits for various stakeholders across the energy system. This fragmentation of DR benefits means that no single agent reaps all the benefits of an investment in DR schemes, potentially making it harder to establish a business model \citep{pinson2014benefits}. Various aspects of the regulatory framework have been described as barriers to DR uptake, with the exclusion of DR from certain revenue streams (e.g. capacity markets) being a frequently mentioned example \citep{lynch2019impacts}. 

For policymakers and stakeholders to reduce barriers to DR, it is imperative that they have detailed information about the competitiveness of different DR technologies. To understand the competitiveness of different consumer flexibility technologies, their costs and benefits have to be estimated and compared, ideally within a framework that also allows their comparison to non-DR flexibility assets like energy storage. 

\begin{table*}[htbp]

\begin{framed}

\nomenclature[]{$r$}{Discount rate}
\nomenclature[]{$\Delta t_{PD}$}{Average plug-in duration in $h$}
\nomenclature[]{$P^{cap}_{app}$}{Application’s required power capacity in $kW$}
\nomenclature[]{$\Delta t_{DD}$}{Application’s discharge duration in $h$}
\nomenclature[]{$N_{app}^{cyc}$}{Application’s number of annual cycles}
\nomenclature[]{$N^{avail}_{V2G}$}{Number of available V2G chargers required}
\nomenclature[]{$N^{contr}_{V2G}$}{Number of contracted V2G chargers required}
\nomenclature[]{$N^{P}_{V2G} $}{Number of required V2G chargers for power}
\nomenclature[]{$N^{E}_{V2G} $}{Number of required V2G chargers for energy}
\nomenclature[]{$N^{adj}_{HP} $}{Number of required HPs adjusted for activation frequency constraints}
\nomenclature[]{$P^{DR}_{t} $}{Available DR power reduction at time $t$ in $kW$}
\nomenclature[56]{$t/T$}{Index/Set of time intervals}
\nomenclature[]{$a_{GMC}$}{Guaranteed minimum charge in $\%$}
\nomenclature[]{$p^{WTA}_{base}$}{Consumers' median willingness-to-accept for base contract of 11.5 hours of RPT in $\$/month$}
\nomenclature[]{$p^{WTA}_{hour}$}{Consumers' median willingness-to-accept per hour of RPT in $\$/month$}
\nomenclature[]{$p^{WTA}$}{Consumers' median willingness-to-accept for DR heating schemes in $\$/month$}
\nomenclature[]{$P^{cap}_{cha}$}{Power capacity of EV charger in $kW$}
\nomenclature[]{$\eta_{cha}$}{Efficiency of charger}
\nomenclature[]{$P_{EV}^{max}$}{EV charger’s maximum power in $kW$}
\nomenclature[]{$P_{bat}^{max}$}{Battery’s maximum power in $kW$}
\nomenclature[]{$E_{app}$}{Application's shifted/discharged energy in $kWh$}
\nomenclature[]{$E_{EV}^{max}$}{EV battery’s energy capacity in $kWh$}
\nomenclature[]{$E_{EV}^{min}$}{EV battery’s minimum energy level in $kWh$}
\nomenclature[]{$E_{EV}^{arr}$}{EV battery’s energy level upon arrival in $kWh$}
\nomenclature[]{$E_{bat}^{max}$}{Battery’s energy capacity in $kWh$}
\nomenclature[]{$E_{bat}^{min}$}{Battery’s minimum energy level in $kWh$}
\nomenclature[]{$t_{EV}^{arr}$}{EV’s arrival time}
\nomenclature[]{$t_{EV}^{dep}$}{EV’s departure time}
\nomenclature[]{$f_{act}^{max}$}{Maximum frequency of activations in heat-pump-only DLC contract per month}
\nomenclature[]{$\Delta T_{b}$}{Maximum permitted temperature divergence in heat-pump-only DLC contract in $K$}
\nomenclature[]{$a_{V2G}^{avail} $}{Availability factor for V2G (fraction of time that an EV is plugged in for)}
\nomenclature[]{$\Delta t^{day}_{cha}$}{Daily time required for charging EV in $h$}
\nomenclature[]{$N_{1D}$}{Number of required unidirectional DR assets}
\nomenclature[]{$P_{1D}^{avg}$}{Average shiftable power consumption of a unidirectional DR asset in $kW$}
\nomenclature[]{$\Delta Q_{b}$}{Building’s permitted heat divergence $kWh$}
\nomenclature[]{$Q_w$}{Thermal energy in hot water storage in  $kWh$}
\nomenclature[]{$P_{1D}^{cap}$}{Power consumption capacity of a unidirectional DR asset in $kW$}
\nomenclature[]{$a_{HC} $}{Fraction of EV’s driving energy charged at home}
\nomenclature[]{$E^{daily}_{drive} $}{Energy used by EV daily for driving in $kWh$}
\nomenclature[]{$t_{act}$}{Time of DR activation}
\nomenclature[]{$COP$}{Heat pump’s coefficient of performance}
\nomenclature[]{$SPF$}{Heat pump’s seasonal performance factor}
\nomenclature[]{$m_b$}{Building’s mass in $kg$}
\nomenclature[]{$c_{p,b}$}{Building’s specific heat capacity in $kJ/(kg\cdot K)$}
\nomenclature[]{$m_w$}{Hot water storage mass in $kg$}
\nomenclature[]{$c_{p,w}$}{Hot water specific heat capacity in $kJ/(kg\cdot K)$}
\nomenclature[]{$\Delta E^{elec}_{HP}$}{HP’s electrical energy needs for heating thermal storage in $kWh$}
\nomenclature[]{$\delta_w$}{Hot water storage density in $kg/m^3$}
\nomenclature[]{$V_{w}$}{Hot water storage volume in $m^3$}
\nomenclature[]{$\Delta T_w$}{Temperature range of hot water storage in $K$}
\nomenclature[]{$L$}{Wall thickness of hot water storage in $m$}
\nomenclature[]{$H$}{Residential ceiling height in $m$}
\nomenclature[]{$A$}{Area rented for hot water storage in $m^2$}
\nomenclature[]{$P_{HP}^{act,avg}$}{Average power use of activated HPs in $kW$}
\nomenclature[]{$\Delta P_{HP}^{red}$}{DR-related reduction in HP power in $kW$}
\nomenclature[]{$\Delta t_{RP}$}{Daily required plug-in time (RPT) in $h$}
\nomenclature[]{$\Delta t_{base}^{plug-in}$}{RPT for $p^{WTA}_{base}$ in $h$}
\nomenclature[]{$p^E_t$}{Electricity spot price in time interval $t$ in $\$/kWh$}

\printnomenclature

\end{framed}

\end{table*}

For cost assessments of electricity generation and storage, capital and operational expenditure are often combined in the levelised cost of electricity (LCOE) or storage (LCOS). These represent the average annuitised cost per unit of electricity (generated or stored). DR, by definition, shifts electricity demand temporally, thereby creating a similar effect on the electricity grid as grid-connected energy storage. As such, a measure for the levelised cost of demand response (LCODR), analogous to the LCOS, is proposed in this paper, providing a measure of the lifetime costs per unit of shifted energy. Two aspects of the lifetime costs of DR have previously been separately estimated: consumer reward payments and the techno-economic costs. Consumer reward payments refer to financial compensation that consumers who participate in a DLC scheme receive and that serves as an incentive for them to enrol. Techno-economic costs refer to all other costs related to a DLC scheme, i.e. investments, operation and maintenance (O\&M) and end-of-life costs of any required infrastructure (e.g. V2G chargers) as well as rebound costs for shifted energy consumption.

Table \ref{tab:lit_table} shows that there is a broad body of literature concerning itself with estimating the consumer reward payments that are needed to enrol consumers in V2G \citep{huang2021electric,parsons2014willingness,lim2016assessment,geske2018willing,kubli2018flexible}, smart charging \citep{bailey2015anticipating,fabianek2024willing,daziano2022willingness,yilmaz2022analysing}, and smart heat pump schemes \citep{kaczmarski2022determinants}. The listed studies exclusively use discrete choice experiments (DCEs), the main method of eliciting stated preferences when revealed preferences cannot be observed (as is often the case with novel technologies like the ones studied here). Other approaches for the estimation of consumer reward payments include contingent valuation, which has been applied by \citet{mehdizadeh2023estimating} to V2G, and the Westendorp price sensitivity meter, which has been applied by \citet{baumgartner2022does} to V2G and by \citet{ensslen2018incentivizing} to smart charging. 

Studies that estimate the techno-economic costs of different DLC schemes are also listed in Table \ref{tab:lit_table}. \citet{al2025impact} assess the costs and benefits of V2G and smart chargers, while \citet{schauble2020conditions} consider the conditions under which smart thermostats are cost-effective investments for residential consumers. \citet{odukomaiya2021addressing} estimate the lifetime hardware costs of thermal storage in residential buildings. Some of the studies in Table \ref{tab:lit_table} also include a framework that allows the costs of DLC schemes and energy storage to be compared. Namely, Vehicle-to-grid (V2G) technology, which is considered a type of DR in this paper, has recently been included in extended LCOS estimations \citep{geng2024assessment,geng4872108techno,rahman2023development}. The recently proposed levelised cost of energy flexibility (LCOEF) \citep{zang2025levelized} is, to the best of the authors' knowledge, the only measure that aims to compare the annuitised lifetime per-unit costs of different DR schemes. However, it also focuses on the techno-economic features and does not consider survey-informed consumer reward payments.

\begin{table}[]
    \centering
    \resizebox{\textwidth}{!}{
    \begin{tabular}{l | c c c c | c c c c}
    \hhline{=========}
         Study & \rotatebox{90}{V2G} & \rotatebox{90}{Smart charging} & \rotatebox{90}{Smart heating/cooling} & \rotatebox{90}{\makecell[tl]{Heat pumps \&\\thermal storage}} & \rotatebox{90}{\makecell[tl]{Consumer reward\\payments}} & \rotatebox{90}{Techno-economic costs} & \rotatebox{90}{\makecell[tl]{Comparison framework\\DLC$\leftrightarrow$Energy storage}} &  \rotatebox{90}{\makecell[tl]{Adjust for\\availability profile}}\\
         \hline
         \citet{yilmaz2022analysing} & & \checkmark & \checkmark & & \checkmark & & \\
         
         \citet{kubli2018flexible} & \checkmark & & \checkmark & & \checkmark & & \\
         \citet{kaczmarski2022determinants} & & & \checkmark & & \checkmark & & \\
         \citet{bailey2015anticipating,fabianek2024willing,daziano2022willingness} &  & \checkmark & & & \checkmark & & \\
         \citet{huang2021electric,parsons2014willingness,lim2016assessment,geske2018willing} & \checkmark & & & & \checkmark \\
         \citet{al2025impact} & \checkmark & \checkmark & & & & \checkmark & \\
         \citet{schauble2020conditions} & & & \checkmark & & & \checkmark & \\
         \citet{odukomaiya2021addressing} & & & & \checkmark & & \checkmark & \\
        \citet{geng2024assessment,geng4872108techno,rahman2023development} & \checkmark & & & & & \checkmark & \checkmark \\
         \citet{zang2025levelized} & \checkmark & & \checkmark & & & \checkmark & \checkmark \\
         \hline
         \textbf{This study} & \textbf{\checkmark} & \textbf{\checkmark} & \textbf{\checkmark} & \textbf{\checkmark} & \textbf{\checkmark} & \textbf{\checkmark}  & \textbf{\checkmark} & \textbf{\checkmark} \\
    \hhline{=========}
    \end{tabular}}
    \caption{Comparison with existing studies}
    \label{tab:lit_table}
\end{table}

Table \ref{tab:lit_table} demonstrates the novelty of this study in three ways: firstly, the framework presented here is the first to integrate consumer payments into a cost measure that allows DLC programs to be compared with energy storage; secondly, this study is the first to combine consumer payments and techno-economic costs into one coherent cost framework; and lastly, the value factor methodology, presented in this work, incorporates the effect of the availability profile of DLC programs, something that existing cost estimates have so far ignored.


The remainder of this paper is organised as follows: Section \ref{sec:LCODR} formally introduces the LCODR as well as the DR schemes and applications for which it is calculated. Section \ref{sec:method} presents the methodology for estimating the LCODR of various pairings of DR schemes and applications, including the estimation of the value factor to adjust for variable availability. Section \ref{sec:data} describes the data that was used for a case study of the levelised DR costs in the context of the British electricity system. Section \ref{sec:res_dis} presents the results, contrasts them with storage costs and discusses their limitations. Section \ref{sec:concl} concludes the paper.

\section{Materials and Methods}

\subsection{Definition}\label{sec:LCODR}

Borrowing from the definition of the LCOS \citep{schmidt2019projecting}, the levelised cost of demand response is the total lifetime cost of a DR scheme per unit of cumulative shifted electricity demand. Shifted electricity is always measured as the delivered reduction in electricity demand rather than the increase in demand at other times. This is in line with the definition of the LCOS, which measures costs per unit of discharged electrical energy \citep{julch2016comparison}. In the case of V2G, shifted electricity can also extend to electricity delivered back to the grid. The computation of the LCODR in energy terms is shown in Equation (\ref{eqb1}).

\begin{equation}
    LCODR\left[\frac{\$}{MWh}\right] = \frac{Investment + \sum^{T}_{t}\frac{O\&M}{(1+r)^t} + \sum^{T}_{t}\frac{Rewards}{(1+r)^t} + 
    \sum^{T}_{t}\frac{Rebound}{(1+r)^t} +
    \frac{End-of-life\; costs}{(1+r)^{T+1}}}{\sum^{T}_{t}\frac{E_{app}}{(1+r)^t}}\label{eqb1}
\end{equation}

For applications that value available power rather than delivered energy, the LCODR in power terms, shown in the supplementary material (S8), may be preferred. The LCODR terms in the numerator are identical to those of the LCOS, except for two differences. Firstly, the charging costs are replaced by rebound costs, which describe the cost of supplying electricity for the shifted electricity demand. Secondly, an additional reward payments term is introduced in Equation (\ref{eqb1}) to reflect the need for financial compensation to consumers.\\

\begin{figure}[t]
\centering
\includegraphics[width=0.5\linewidth]{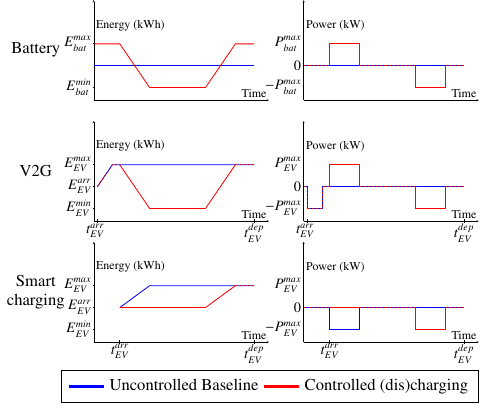} 
\caption{Energy and power profiles for servicing an energy storage application with different technologies. Note that the minimum energy level for EVs ($E_{EV}^{min}$) is their energy capacity ($E_{EV}^{max}$) multiplied by the guaranteed minimum charge ($f_{GMC}$): $E^{min}_{EV} = E^{max}_{EV} \times f_{GMC}$}
\label{fig:ex}
\end{figure}

Figure~\ref{fig:ex} shows how a single discharge cycle can be provided using a battery, a V2G-capable vehicle, or a unidirectional smart charger. Storage and DR assets can be used to service a variety of applications, such as energy arbitrage or various ancillary services. The LCOS is normally estimated individually for each potential application. Any application that is centred on active power delivery, can be characterised by four main features: response time, rated power, discharge duration, and the annual number of discharges \citep{schmidt2019projecting}. Response time is a prerequisite for a technology to be able to service a certain application. The other three features strongly impact the required design of a storage or DR asset, and thereby greatly influence the resulting levelised costs. 

The power and energy capacities of storage technologies can be designed relatively independently to meet the characteristics of a specific application. The domestic DR assets that are considered in this paper, on the other hand, are often much more constrained in their power-to-energy capacity ratio. For V2G and smart charging, for example, the energy capacity is very difficult to change for a single vehicle/charger, because this would imply redesigning the EV's battery. The power capacity of the charger would be easier to alter but is usually limited by the power rating of a household's grid connection. Charger power and the energy capacity of an EV's battery are therefore considered unchangeable in the rest of this paper. The only DR scheme where the power-to-energy capacity ratio is not considered set is the combination of heat pump and hot water tank (HP + thermal storage), where the tank's volume is independent of the power capacity.

With DR's fragmented business case, different elements of the costs are incurred by different stakeholders, meaning it is important to specify the perspective of the cost assessment. The proposed LCODR approach takes the perspective of the beneficiary from responsive consumers, rather than the consumers themselves. In recent literature, this beneficiary is often referred to as an "aggregator" which could be an energy supplier or an intermediary that sells flexibility from DR to grid operators, generators or suppliers. What actors can take the role of beneficiary/aggregator depends on the policy context. In many current electricity systems, this role can only be taken by the energy supplier or the regional utility. The proposed framework is policy-agnostic, and may instead serve as a basis for evaluating future regulation. The proposed DR schemes all fall under the DR category of direct load control (DLC), meaning they do not require consumers to directly engage with a price signal. Instead, consumers are paid monthly compensation for giving away some control over one of their assets. In the assessed DR scenarios,  installing the required infrastructure is considered a cost to the aggregator, meaning consumers have it installed at no extra cost to them as part of their DR contract.

\subsection{Scope}

\subsubsection{Demand response schemes and applications}

\begin{figure}
    \centering
    \includegraphics[width=0.5\linewidth]{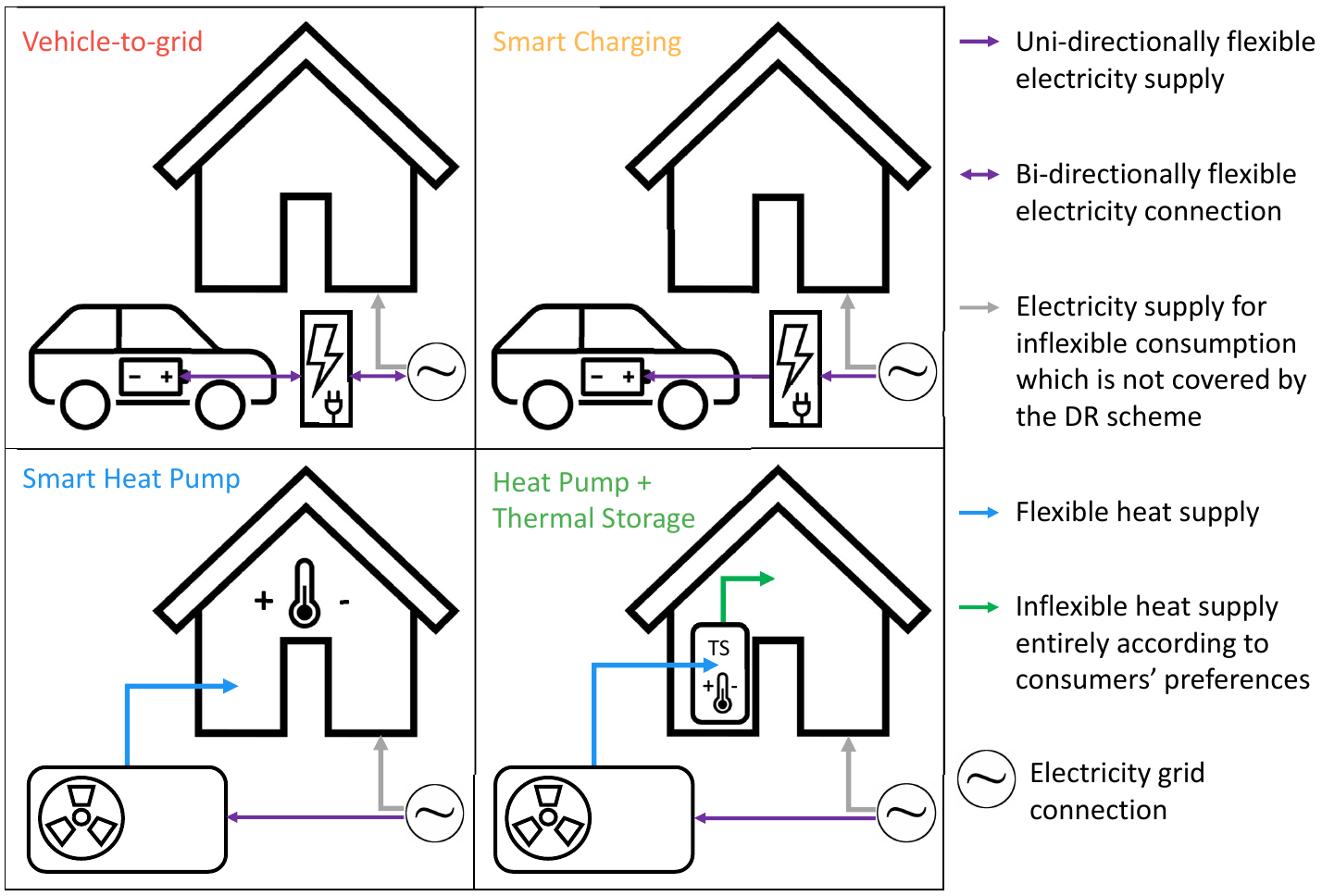} 
    \caption{Illustration of the components and estimation process of the availability profile-adjusted LCODR ($LCODR_{VF}$)}
    \label{fig:illu_schm}
\end{figure}

The levelised costs for four different DR schemes are considered in this paper: V2G, EV smart charging, smart heat pumps and heat pumps with thermal storage. They are illustrated in Figure~\ref{fig:illu_schm}  and their details are listed below.\begin{enumerate}[leftmargin=*]
    \item  \textbf{V2G} describes the delivery of power from EVs to the grid and requires the installation of a bidirectional charger. Consumers on this contract indicate their departure time at which the aggregator ensures a fully charged battery. Before departure, the aggregator is free to discharge the battery to a specified guaranteed minimum charge (GMC), which the consumer reserves for unexpected journeys. Consumers commit to fulfilling a daily required plug-in time (RPT) and receive a monthly reward for making their vehicle available for V2G. 
    \item  \textbf{Smart charging} contracts work much in the same way as V2G, except that they work with unidirectional chargers. Rather than discharging, smart charging allows the aggregator to shift the charging time within the plugin period (again while ensuring a full battery at the indicated departure).
    \item Consumers on the \textbf{smart heat pump} scheme specify a maximum temperature setting divergence ($\Delta T_b$) which allows the aggregator to temporarily increase their house's temperature above or below the specified setting, thereby shifting electricity consumption forward or backwards, respectively. The aggregator can only do this for a specified maximum number of activations per month ($f_{act}^{max}$). 
    \item The  \textbf{heat pump with thermal storage} tariff installs a heat storage tank within the consumer's home which will be entirely controlled by the aggregator while ensuring the heating preferences of the consumer are met without exception. This scheme differs from the rest in that it potentially does not impact consumers' lives at all (not even having to indicate preferences like departure times). For that reason, no conventional reward payments are considered for this tariff. Instead, consumers are compensated for the space that the hot water tank takes up (the installation of which is already paid for by the aggregator).
\end{enumerate}

The 12 storage applications from \citep{schmidt2019projecting} are considered in this paper as potential applications for DR. Table \ref{tab:my_label} shows the suitability of the DR schemes for the 12 applications. Response times should not be an issue for any of the DR technologies, with \citet{Zhang2022} and \citet{lee2020providing} demonstrating that even primary response can be achieved with heat pumps, and \citet{o2022frequency} and \citet{falahati2016new} showing the same for V2G and smart charging, respectively.
Black start, power reliability and power quality services require active power delivery rather than just load shifting, so unidirectional DR assets are not suitable for this. Seasonal storage requires long discharge durations that cannot be serviced by any of the DR assets.
\begin{table}[htbp]
    \centering
    \resizebox{0.5\linewidth}{!}{%
    \begin{tabular}{l|c c c | c c c c}
         \hhline{========}
          & \makecell[tl]{$P_{app}^{cap}$ \\ MW}& \makecell[tl]{$\Delta t_{DD}$ \\ h} & $N^{cyc}_{app}$ & \makecell[tl]{V2G} & \makecell[tl]{Smart\\charging} & \makecell[tl]{Heat\\pump} & \makecell[tl]{HP \& \\thermal\\storage}\\
         \hline
         Energy arbitrage & 100 & 4 &300&\checkmark & \checkmark & \checkmark & \checkmark \\
         Primary response & 10 & 0.5 &5000& \checkmark & \checkmark & \checkmark & \checkmark \\
         Secondary response & 100 & 1 &1000& \checkmark & \checkmark & \checkmark & \checkmark \\
         Tertiary response & 100 & 4& 10 & \checkmark & \checkmark & \checkmark & \checkmark \\
         Peaker replacement & 100 &4& 50& \checkmark & \checkmark & \checkmark & \checkmark \\
         Black start & 10 & 1 & 10 &\checkmark & & & \\
         Seasonal storage & 100 & 700 & 3 & & & \\
         T\&D invest. deferral & 100 & 8 & 300 & \checkmark & \checkmark & \checkmark & \checkmark \\
         Congest. management &100& 1& 300 & \checkmark & \checkmark & \checkmark & \checkmark \\
         Bill management &1& 4 & 500 & \checkmark & \checkmark & \checkmark & \checkmark \\
         Power quality &1&0.5& 100& \checkmark &  & & \\
         Power reliability &1&8& 50&\checkmark &  &  &  \\
         \hhline{========}
    \end{tabular}}
    \caption{Suitability of DR schemes for storage applications including application requirements from \citep{schmidt2019projecting}}
    \label{tab:my_label}
\end{table}

\subsubsection{Geographic and socio-demographic applicability}

The LCODR framework can only be applied in contexts (hypothetical or real) where individual transportation and/or spatial heating/cooling are amply electrified. It is furthermore necessary that spatial heating/cooling and individual transport options are sufficiently widespread within a studied region. This may not be the case for economic or climatic reasons. The LCODR framework was developed in the context of the United Kingdom (UK), a developed Western country with a temperate, maritime climate, including cold, wet winters and warmer, drier summers \citep{kottek2006world}. It is further implied that EVs and heat pumps reach mainstream adoption, as is foreseen by current net-zero plans \citep{burnett2024uk}. These conditions should be considered when applying the framework to other geographic contexts. Conversely to the methodological framework, the results from this study cannot be interpreted outside their context, because they depend a lot on the input data. The equipment cost data is sourced from the UK, while DCE data is based on surveys in countries with generally similar income levels to the UK.The representativeness of the DCE studies is discussed in the supplementary material (S7), concluding that the samples slightly overrepresent men and are strongly skewed towards high-income countries. This reinforces the results' applicability limit to countries with similar income levels to the UK.

\subsection{Estimation model }\label{sec:method}

\begin{figure}
    \centering
    \includegraphics[width=0.5\linewidth]{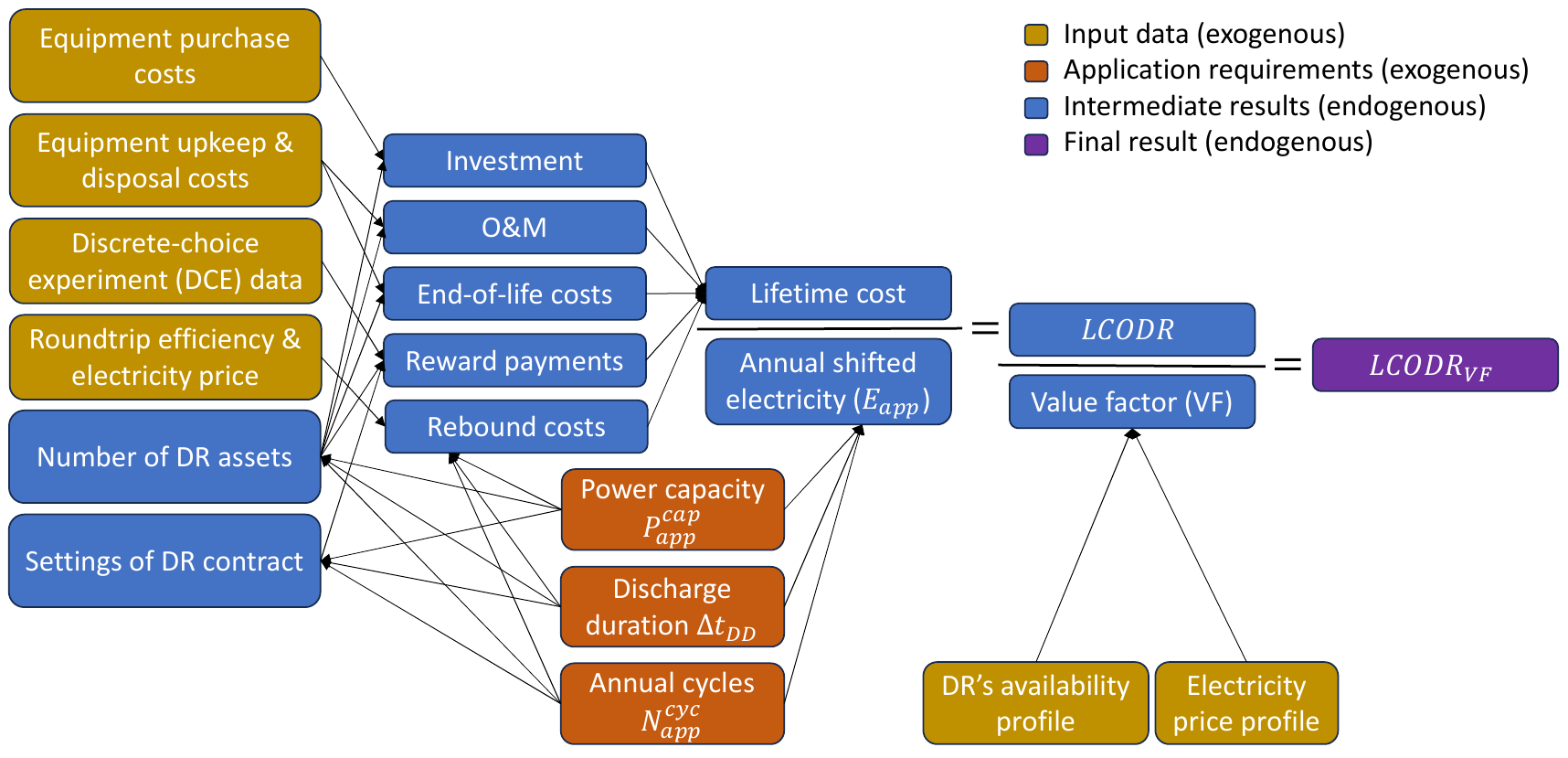}
    \caption{Illustration of the components and estimation process of the availability profile-adjusted LCODR ($LCODR_{VF}$)}
    \label{fig:illu_est}
\end{figure}

In the following, a consumer who is enrolled in a DLC scheme and has the required infrastructure installed will be labelled a "DR asset". To identify the cost of delivering a certain service, it is crucial to identify the number of DR assets that are required to be available to deliver said service (Section~\ref{num_ass}). The number of assets and their reward payments can further be affected by the required discharge duration (Section~\ref{dd_cons}) and the number of annual activations (Section~\ref{cons_nc}). To adjust the LCODR for differences in availability profile, it is divided by the value factor (Section~\ref{Availability adjustments}). To account for uncertainties in all of the input data, a Monte-Carlo simulation is carried out (Section~\ref{mc_sim}). Figure~\ref{fig:illu_est} illustrates the estimation process for the availability profile-adjusted LCODR. It differentiates input parameters and intermediate variables. Input parameters are either from the data sources detailed in Section~\ref{sec:data} or dictated by the requirements of the application for which the LCODR is being estimated (see Table~\ref{tab:my_label}).

\subsubsection{Estimating the number of required demand response assets} \label{num_ass}
 The methodology for identifying the number of required assets varies between V2G and unidirectional DR schemes.

Given that the chargers' power capacity and the EVs' energy capacity are considered immutable, the number of available chargers required for V2G, $N^{avail}_{V2G}$, is either dictated by the power, $P_{app}^{cap}$, or the energy, $E_{app}$, requirements, depending on the application. To make sure that both energy and power requirements are met, the higher number between the energy-mandated ($N_{V2G}^{E}$) and power-mandated ($N_{V2G}^{P}$) number of chargers has to be chosen.

\begin{equation}
    N^{avail}_{cha} = max(N_{cha}^{P},N_{cha}^{E}) \label{eq2}
\end{equation}

The number of available V2G chargers required to provide the needed power ($N_{V2G}^{P}$) is simply the power capacity required by the application ($P_{app}^{cap}$) divided by the effective power capacity of an individual bidirectional charger($P^{cap}_{cha} \times \eta_{cha}$).

\begin{equation}
    N_{V2G}^{P} = \frac{P^{cap}_{app}}{P^{cap}_{cha} \times \eta_{cha}} \label{eq3}
\end{equation}

Similarly, the number of available V2G chargers needed to provide the required energy capacity ($N_{V2G}^{E}$) is obtained by dividing an application's required energy capacity ($E_{app}^{cap}$) by the available energy capacity of a singular EV ($E_{EV}^{max} - E_{EV}^{min}$). The available energy capacity of a single EV is dictated by its battery's capacity ($E_{EV}^{max}$) and the guaranteed minimum charge ($a_{GMC}$) which is the minimum level of energy that a V2G contract commits to always leaving in a vehicle's battery. The needed energy capacity for an application is the product of power capacity ($P_{app}^{cap}$) and discharge duration ($\Delta t_{DD}$).

\begin{equation}
    N_{V2G}^{E} = \frac{E_{app}}{E_{EV}^{max} - E_{EV}^{min}} = \frac{P_{app}^{cap} \times \Delta t_{DD}}{E_{EV}^{max} \times (1-a_{GMC})}
\end{equation}

The estimations above are for the required number of available chargers. A key difference between DR and storage is that DR assets are not always available (e.g. V2G-capable EVs are not always plugged in). To obtain a cost estimate, it is crucial to obtain the number of contracted chargers that are required to ensure that, on average, there are enough available chargers. The number of available chargers is a product of the number of contracted chargers and the availability factor ($a_{V2G}^{avail}$), which is the fraction of time that each charger is available for. This relationship can be rearranged to calculate the number of contracted chargers required

\begin{equation}
    N^{contr}_{V2G} = \frac{N^{avail}_{V2G}}{a_{V2G}^{avail}}
\end{equation}

The available time per day for bidirectional charging is simply the average daily time that the vehicle is plugged in for, and not charging. The amount of time that EVs will be plugged in for, if drivers get paid for plugging in, is difficult to estimate because few such schemes exist today. Many V2G choice experiments \citep{huang2021electric,lee2020willingness} include a minimum daily RPT term, though. It is assumed that consumers plug in for as long as they are required by their contracts ($\Delta t_{RP}$), and no longer. The daily charging time ($\Delta t^{day}_{cha}$) can be determined from the effective charger power ($P^{cap}_{cha} \times \eta_{cha}$) and the average daily energy used for driving ($E^{daily}_{drive}$). $f_{hc}$ is the fraction of the latter which is charged at the domestic V2G charger.

\begin{equation}
    a_{V2G}^{avail} = \frac{\Delta t_{RP} - \Delta t^{day}_{cha}}{24h} = \frac{\Delta t_{RP} - \frac{E_{drive}^{day} \times a_{HC}}{\eta_{cha} \times P_{cha}^{cap}}}{24h} \label{avail_fac}
\end{equation}



Dividing by the availability factors only means that the required capacity will be available on average. While this ensures that, on average, the DR assets can provide the same service as storage, their profile may vary quite significantly from the flat profile of a continuously available storage asset. Section~\ref{Availability adjustments} introduces a methodology to adjust for an uneven availability profile.

In contrast to V2G, unidirectional DR schemes have the number of required available assets dictated only by power requirements. This is because the available power is dictated by the uncontrolled power consumption that can be moved by DR. This consumption also dictates the energy that can be moved, meaning that meeting the power requirements also ensures that the energy requirements are met.
Equation (\ref{eq5}) gives the number of unidirectional assets ($N_{1D}$) required to fulfil the power requirements, which is similar to that of V2G from Equation (\ref{eq3}).

\begin{equation}
    N_{1D} = \frac{P_{app}^{cap}}{P_{1D}^{avg}} \label{eq5}
\end{equation}

Unidirectional DR assets are only available when they would otherwise be consuming in an uncontrolled scenario, meaning usually more need to be contracted than in the V2G scenario to ensure enough of them are available. No availability factor has to be applied to Equation (\ref{eq5}) because utilising the average power is equivalent to adjusting the power capacity for its availability (i.e. $P_{1D}^{avg} = P_{1D}^{cap} \times a_{1D}^{avail}$).






\subsubsection{Discharge duration constraints}  \label{dd_cons}
\begin{figure}[t]
\centering
\includegraphics[width=0.5\linewidth]{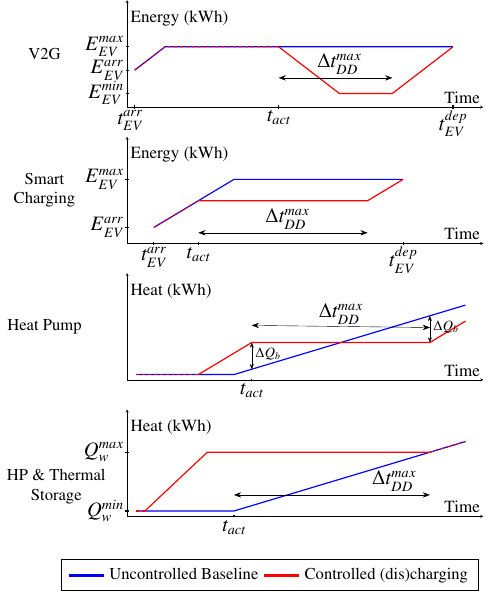}
\caption{Discharge duration limits illustration on the energy profiles of different DR assets. Note that the heat pump-only profile (3\textsuperscript{rd} graph) shows an activation that reduces the heat pump's power consumption to zero ($ \Delta P^{red}_{HP} = P^{act,avg}_{HP}$)}
\label{fig:exDD}
\end{figure}

Even though unidirectional DR assets do not discharge energy, the term discharge duration will be used throughout this paper to denote the duration of activation, i.e. duration of lowered electricity consumption. The discharge duration is limited by the rebound for unidirectional DR assets (recharge for V2G). This is because the energy that is required to fulfil the rebound requirements would count against the DR activation and thereby annul it, even if other assets were continuing the demand reduction. Figure~\ref{fig:exDD} shows how the rebound limits the discharge duration for the different DR assets.

For both EV-based DR schemes, the discharge duration depends on the average duration for which vehicles are plugged in. While data exists for current plug-in durations \citep{dftEVdata}, this does not take into account that plug-in durations will likely be longer when consumers are financially incentivised to plug in for longer. Consumer survey data for the effect of financial incentives on daily RPT is available both for V2G (e.g. \citep{huang2021electric,parsons2014willingness,lim2016assessment}) and smart charging \citep{fabianek2024willing}. RPT states the number of hours per day that an EV has to be plugged in for and can only be translated into the average plug-in duration by assuming a plug-in frequency. An EV that has an average daily plug-in time of 20 hours, for example, will have an average plug-in duration of 10 hours if its plug-in frequency is twice per day. In the following, an average plug-in frequency of once per day is assumed, resulting in the average plug-in duration being equal to the daily RPT ($\Delta t_{PD} =\Delta t_{RP}$). \\

For V2G, the discharge may start at any point during the time that an EV is available for V2G services, which will be simulated by assuming that discharge starts halfway through the plug-in period. This means that, during actual operation, the discharge duration may be constrained to be shorter or allowed to be longer, but on average it will be constrained by Equation (\ref{eq9}). The time needed for charging driving energy needs ($E^{daily}_{drive} \times a_{HC}$) and recharging any V2G-discharged energy ($E^{max}_{EV} - E^{min}_{EV}$) cannot be used for V2G, and therefore has to be subtracted.

\begin{equation}
    \Delta t_{DD}^{max} = \frac{\Delta t_{PD} - \frac{E^{daily}_{drive} \times a_{HC}}{\eta_{cha}\times P^{cap}_{cha}}}{2} - \frac{E^{max}_{EV} - E^{min}_{EV}}{\eta_{cha}\times P^{cap}_{cha}} \label{eq9}
\end{equation}

The activation of flexibility from smart charging can only occur when the EV would otherwise be charging, which, in an uncontrolled charging scenario, is only the case for a limited time window after plugging in. Equation~(\ref{dd_sc}) shows how the remaining time therefore acts as an upper constraint for the discharge duration, because afterwards, the previously delayed energy will have to be charged. Note that several EVs are needed to achieve a discharge duration that is longer than the uncontrolled charging time. Once an EV is forced to recharge because its departure is near, however, the recharge would work against the effect of the next EV's charging delay, turning the plug-out time into an upper limit for the discharge duration.

\begin{align} 
    \Delta t_{DD}^{max} = \Delta t_{PD} - \frac{E^{daily}_{drive} \times a_{HC}}{P_{cha}^{cap} \times \eta_{cha}} \label{dd_sc}
\end{align}

For the heat pump-only scheme, the discharge duration is limited by the allowed temperature divergence ($\Delta T_b$) and the building's heat capacity ($m_b c_{p,b}$), which is the amount of thermal energy it takes to heat the building by one degree Celsius. Together, these dictate the possible difference in the building's thermal energy ($\Delta Q_b$), which can be used to deduce the discharge duration as shown in Figure~\ref{fig:exDD}. Note that Figure~\ref{fig:exDD} shows a case in which the heat pump is used to actively heat up the building, but if the heat pump only maintained the temperature to make up for heat losses, the inferred relationships would remain the same. The difference in the red and blue slopes is the reduction in the effective heating rate, which is the product of the reduction in heating power and the seasonal performance factor ($\Delta P_{HP}^{red} \times SPF$). The $SPF$ is an annual power-weighted average of the Coefficient of Performance ($COP$), which measures the heat output of a heat pump per unit of electrical energy input. Dividing the difference in electrical energy input ($E^{elec}_{HP}$) by the reduction in heating power ($\Delta P_{HP}^{red}$) gives the duration associated with a certain temperature threshold ($\Delta T_b$). This duration is doubled to obtain the maximum discharge duration ($\Delta t^{max}_{dd}$) because the temperature setting can be exceeded by $\Delta T_b$ before activation and subsided by $\Delta T_b$ afterwards.

\begin{equation}
    \Delta t^{max}_{DD} = 2\times \frac{\Delta E^{elec}_{HP}}{\Delta P_{HP}^{red}} = 2\times \frac{\Delta Q_b}{\Delta P_{HP}^{red}\times {SPF}} = 2 \times \frac{m_b c_{p,b} \Delta T_b}{\Delta P_{HP}^{red} \times {SPF}}
\end{equation}
 
The possible reduction in heating power decreases as discharge duration goes up because longer discharge durations require longer reductions in heating power, which can only remain within the tolerated temperature divergence ($\Delta T_b$) if the heating power is reduced by less. It would be impossible for the power reduction to exceed the heat pump's active power consumption which is accounted for in Equation (\ref{eqa}). $P_{HP}^{act,avg}$ denotes the average power consumption of activated heat pumps, i.e. the average power consumption when a heat pump is turned on. Even though heat pumps are a unidirectional DR asset, $P_{HP}^{act,avg} \neq P_{1D}^{avg}$ because the former denotes the average non-zero power whereas the latter includes times when the heat pump is off.

\begin{equation}
\Delta P_{HP}^{red} \leq P^{act,avg}_{HP} \label{eqa}
\end{equation}

For the tariff that includes thermal storage, the discharge duration is limited by how long the thermal storage can sustain the heat supply that would otherwise have been supplied by the average power consumption of the heat pump when switched on ($P^{act,avg}_{HP}$). Since the heat storage can be designed to always meet the entire heating demand, a DR activation in this scheme always reduces the power demand to zero ($\Delta P_{HP}^{red} = P^{act,avg}_{HP}$), to minimise the number of required assets under contract.

\begin{equation}
    \Delta t_{DD}^{max} = \frac{\Delta E^{elec}_{HP}}{\Delta P_{HP}^{red}} = \frac{Q_{w}^{max} - Q_{w}^{min}}{P^{act,avg}_{HP} \times {SPF}} = \frac{m_{w}c_{p,w}\Delta T_w^{max}}{P^{act,avg}_{HP} \times SPF}
\end{equation}

The mass of water within the thermal storage can be calculated from its density and the volume of the storage. It is assumed that the hot water storage tanks are cylindrical, but that reward payments have to be paid for the associated square area on which the cylinder sits. Equation \ref{eq13} expresses water mass as a function of the rented area ($A$).

\begin{equation}
    m^w = \delta_wV_{st} = \delta^w((H - 2L)\times (\pi \times (\frac{\sqrt{A}}{2} - L)^2))\label{eq13} 
\end{equation}

Here, $H$ is the ceiling height and $L$ is the insulation wall thickness.


\subsubsection{Constraints for the number of cycles}  \label{cons_nc}
The smart heat pump scheme (without thermal storage) is the only one that limits the number of annual cycles. This is because an activation affects consumers directly as it influences the temperature within their home. The allowed number of weekly activations is a key contract term for the heat pump scheme, and if there are more cycles than can be provided by a single contracted heat pump, more heat pumps are needed following the relation shown in Equation (\ref{eq14}).

\begin{equation}
    N^{adj}_{HP} = \frac{f_{act}^{max}\times 12}{N_{app}^{cyc}} \times N_{1D}  \label{eq14}
\end{equation}

Here, $N^{adj}_{HP} $ denotes the number of contracted heat pumps once adjusted for the number of allowed activations. This is expressed as a function of the unadjusted number of heat pumps ($N_{1D}$), the maximum number of activations per month ($f_{act}^{max}$), and the annual cycle requirement of the storage application ($N_{app}^{cyc}$).

\subsubsection{Value factor} \label{Availability adjustments}
The value of electricity is strongly influenced by the time at which it is made available. When assessing the cost of VRE, this issue was quickly realised and addressed by a broad body of literature \citep{simpson2020cost,loth2022we,hirth2013market,joskow2011comparing,ueckerdt2013system}. With LCOS being equivalent to LCOE for storage technologies, DR assets can be considered to behave analogously to VRE generation in that they vary in their availability. The methods that adjust for the variability of VRE can therefore be applied to address the variable availability of DR. In this paper, the value factor from \citep{hirth2013market} is adapted. To obtain this for a given VRE asset, its generation profile is multiplied by the energy price profile and then divided by the same product for a flat generation profile. The same procedure is applied for DR, with the main difference being that rather than generated electricity, the potential for electricity demand reduction (or discharge in the case of V2G) is measured, denoted in Equation (\ref{eq15}) as $P^{DR}_t$.

\begin{equation}
    VF = \frac{\sum_{t=1}^{t=T}p_t^EP^{DR}_t}{\sum_{t=1}^{t=T}p_t^E\frac{\sum_{t=1}^{t=T}P^{DR}_t}{T}} \label{eq15}
\end{equation}


For unidirectional DR assets, $P^{DR}_t$ is simply the power that would be consumed in an uncontrolled scenario in time period $t$ and can be moved away. For V2G, $P^{DR}_t$ is the available charger power that can be discharged in time period $t$. When the energy requirements dictate the number of V2G chargers (i.e. $N_{cha}^E > N_{cha}^P$), $P^{DR}_t$ has to be replaced by $E^{DR}_t$, which is the energy available for discharging in time period $t$. To assess the power and energy profiles for V2G, the virtual battery from \citet{thran2024reserve} was used, with $P^{DR}_t$ being inferred from the power boundary and $E^{DR}_t$ being the difference between the energy boundaries. A value-adjusted LCODR ($LCODR_{VF}$) can be computed by dividing the original cost estimate by the value factor as shown in Equation (\ref{eq16}).

\begin{equation}
    LCODR_{VF} = \frac{LCODR}{VF} \label{eq16}
\end{equation}

Akin to a solar panel, which can only produce electricity when the sun shines, an EV can only provide V2G services when it is plugged into a bidirectional charger. This means that its availability profile is dictated by the consumer's plug-in behaviour, which is assumed to be exogenous (like the weather in the case of the panel). It is important to note here that the availability factor ($a^{V2G}_f$) from Equation (\ref{avail_fac}) denotes what proportion of the time the EV will be available for V2G (equivalent to a capacity factor for solar). The value factor only denotes the relative value of times at which an asset is available. For unidirectional DR, the availability factor is also considered exogenous, whereas for V2G it is a function of the contract terms (i.e., the required plug-in time $\Delta t_{RP}$).

Other methods exist for time-weighting the value of VRE, which were also considered for the LCODR. The levelised cost of avoided electricity (LACE) employs a similar methodology to the value factor, but additionally includes adjustments for potential capacity payments \citep{eia2013levelized}. This would be difficult to estimate for the LCODR because the capacity credit (rating that reflects the ability to provide system reliability reserves) for DR assets is uncertain, as they have so far mostly been excluded from capacity markets \citep{lynch2019impacts}. \citet{joskow2011comparing} proposes evaluating VRE assets both on their costs as well as their potential profits. While this is a very precise method, it does not generalise well, meaning it will not work well with a general measure like the LCODR. Lastly, the system LCOE from \citet{ueckerdt2013system} analyses system factors affecting the value of VRE, distinguishing between grid costs, balancing costs, and profile costs. Using an equivalent of their comprehensive approach in the cost-comparison of demand response and energy storage may be of interest for future research, but it goes beyond the scope of this study. Balancing costs for VRE arise from its unpredictability. While unpredictability is a factor that sets apart DR from energy storage, several studies (e.g. \citep{thran2024reserve, marevcek2023predictability}) show that for sufficiently large DR asset ensembles, aggregate availability and device status become increasingly predictable.

\subsubsection{Monte-Carlo simulation}  \label{mc_sim}
To reflect the uncertainty of the input parameters that are presented in the following section, a Monte Carlo simulation is implemented. The methodology for this largely follows that of \citet{schmidt2024monetizing} with the number of drawn samples set to 1000. Monte Carlo samples are taken from a normal distribution that is truncated at 1.285 standard deviations above and below the mean input value. The mean input values for different parameters are given in the various tables in Section~\ref{sec:data}. Informed by the spread of parameter estimates where several sources were consulted, the standard deviations are assumed to be 33\% for all inputs. For the estimated value factors, a standard deviation of 10\% is assumed to reflect the possible effect of increased VRE generation or a potential price cannibalism effect caused by higher penetrations of storage or DR. The uncertainty in the value factors is expected to be relatively low because some established electricity price profile patterns (e.g. expensive evening hours) would have to significantly change for them to vary. Usage patterns may also affect the value factor, the sensitivity to which is investigated by running a separate Monte Carlo simulation for value factors, where a subset of 50 DR assets is simulated 1000 times.

\subsection{Data}\label{sec:data}
Differences in socio-economic patterns, car usage, and climate mean that the LCODR is better computed for specific regions rather than globally. The presented data in this section is for a case study of the LCODR in Great Britain. Because of the low prevalence of domestic DLC schemes, data sources are not always available for Britain so data was also taken from studies that were conducted in regions with similar socio-economic structures. In general, the lack of accessible DLC consumer engagement data has meant that data sometimes had to be repurposed out of context, reducing the robustness of the results. 

\subsubsection{Investment costs}
Capital costs for the different DR schemes are given in Table~\ref{tab:inv_costs}. Note that all investment costs measure the upgrade from an existing domestic charging/heating set-up, as consumers are expected to install this regardless of DR participation. The first three  DR schemes all require investment in only one piece of infrastructure (first three rows of Table~\ref{tab:inv_costs}). Heat pumps with thermal storage, however, require both a smart thermostat and hot water storage. Advanced metering infrastructure is also a prerequisite for any of the DR tariffs and can come at a considerable cost \citep{kaczmarski2022determinants}. However, many countries, including the United Kingdom, have pledged to roll out smart meters, meaning this will likely not be a cost to the DR aggregator \citep{smartmeter} and is therefore not considered in this study.

\begin{table}[htbp]
    \centering
    \resizebox{0.5\linewidth}{!}{%
    \begin{tabular}{l l l l}
        \hhline{====}
         Parameter & Unit & Value & Source \\
         \hline
         V2G charger cost & \$/charger (7.4kW) & 3,000 & \citep{geng2024assessment,geng4872108techno} \\
         Smart charger cost & \$/charger (7.4kW) & 107 & \citep{heilmann2021much} \\
         Smart thermostat & \$/thermostat & 85 & \citep{schauble2020conditions} \\
         Hot water storage & \$/m\textsuperscript{3} & 2,042 & \citep{odukomaiya2021addressing}\\ 
         \hhline{====}
    \end{tabular}}
    \caption{Investment costs}
    \label{tab:inv_costs}
\end{table}

\subsubsection{Reward payments}
Accurately estimating reward payments is a difficult aspect of the LCODR data collection. Few commercial DR schemes exist for the four variants considered in this paper, so consumers' preferences cannot be inferred from them, especially when considering their choices regarding specific contract settings like RPT for the EV-based schemes. When consumers' revealed preferences cannot be observed by looking at their purchasing habits (i.e. enrolment data for existing commercial DR schemes), consumers' stated preferences can instead be captured through discrete-choice experiments (DCEs) \citep{samuelson1948consumption}. DCEs ask participants to choose between hypothetical alternative scenarios of a good or service with different attributes. The relative importance of these attributes can then be inferred from their choices \citep{mangham2009or} and can be translated into the willingness-to-pay (WTP) for an attribute. For DLC schemes, where the aggregator pays the consumers, the equivalent measure for WTP is the willingness-to-accept (WTA), which states the amount of monthly financial reward participants associate with the terms of a DR contract.

Depending on their design (forced or unforced choice), DCEs can estimate only the WTA for a specific attribute of a DR contract (e.g. GMC or RPT) or also include the base WTA for this type of DR contract in general. For the EV schemes, the most relevant attribute is the RPT, for which several studies estimate the WTA ($p^{WTA}_{hour}$) as shown in Table \ref{tab:rewards}. For the base WTA ($p_{base}^{WTA}$), only a single study was found for V2G (\citep{huang2021electric}) and smart charging (\citep{daziano2022willingness}), respectively. To increase the robustness of these estimations, they were compared to existing V2G\footnote{Octopus Energy - Power Pack: \url{https://octopus.energy/power-pack/} [Accessed: 2024-09-05]} and smart charging\footnote{OVO Energy - Charge Anytime: \url{https://www.ovoenergy.com/electric-cars/charge-anytime}, EdF Energy - EV tariffs: \url{https://www.edfenergy.com/electric-cars/ev-tariffs}, Octopus Energy - EV tariffs: \url{https://octopusev.com/ev-hub/guide-to-energy-tariffs-for-electric-cars} [All accessed: 2024-09-05]} schemes, confirming the range of their estimations. The reward payments per charger for a given RPT can be calculated by using the base cost and adding the WTA per hour of RPT multiplied by the additional number of hours. For an exemplary smart charging contract with an RPT of 15 hours, the reward payment estimation is shown in Equation (\ref{eq17}).

\begin{equation}
    Rewards = p^{WTA}_{base} + (\Delta t_{RP} - \Delta t^{plug-in}_{base}) \times p^{WTA}_{hour} = 40.81 + (15 - 10) \times 11.8 = 99.81 \$/Charger
    \label{eq17}
\end{equation}

To ensure that Equation (\ref{eq17}) does not produce negative reward payments, a minimum monthly reward of \$5 was assumed for V2G, smart charging and heat pumps with thermal storage. Regarding reward payments for the heat pump tariff, data from a controlled air conditioning DCE in the US is repurposed under the assumption that temperature differences cause the same discomfort, be it for cooling or heating \citep{kaczmarski2022determinants}. The maximum temperature divergence ($\Delta T_b$) in that trial was 1.67\degree Celsius, and the maximum activation frequency ($f_{act}^{max}$) was 3.33/month. For the thermal storage scheme, the WTA for the occupied space was estimated from average rent and living space data. The procedure for estimating the WTA for all schemes is explained in more detail in the supplementary material (S1 - S5). All DCE data represents the median WTA, i.e. the reward payment at which half of the sample population is expected to participate in a DR scheme. At low penetrations of DR, WTA may be significantly cheaper because consumers with a lower WTA (e.g. early adopters) could be targeted first.

\begin{table}[htbp]
    \centering
    \resizebox{0.5\linewidth}{!}{%
    \begin{tabular}{l l l l l}
        \hhline{=====}
         DR Scheme & Measure & Unit & Value  & Source \\
         \hline
         V2G & $p^{WTA}_{base}$ &\$/month/charger & 59.1 & \citep{huang2021electric}\\ 
          & $p^{WTA}_{hour}$ & \$/month/hour(RPT) & 29 & \citep{huang2021electric,parsons2014willingness,lim2016assessment}\\
         Smart charging & $p^{WTA}_{base}$ & \$/month/charger & 40.8 & \citep{daziano2022willingness}\\ 
          & $p^{WTA}_{hour}$ & \$/month/hour(RPT) & 11.8 & \citep{fabianek2024willing, daziano2022willingness}\\ 
         Heat pump & $p^{WTA}$ &\$/month/thermostat & 10.7 &\citep{kaczmarski2022determinants}\\
         HP \& storage & $p^{WTA}$ &\$/month/m\textsuperscript{2} & 17.7 & \citep{housing2024,team_2024}\\
         \hhline{=====}
    \end{tabular}}
    \caption{Reward payments}
    \label{tab:rewards}
\end{table}

\subsubsection{ Miscellaneous data} \label{vfdata}

In line with recent studies \citep{huber2021vehicle}, the annual operation and maintenance costs ($O\&M$) are assumed to be 5\% of the capital investment and the lifetime is assumed to be 15 years. End-of-life costs are assumed to be zero for the smart charging and smart heat pump tariff, 50\$ per V2G charger, and 10\$ per square meter of hot water tank surface.

The estimates for additional parameters, including their sources, are shown in Table~\ref{tab:tech_params}. Note that, the temperature difference for the hot water storage ($\Delta T_w$) can take any value, as long as the high temperature does not exceed 100\degree C. A higher value of $\Delta T_w$ results in higher losses from decreased COPs and heat dissipation while also decreasing the required storage volume and thereby area. Optimisation could be applied to $\Delta T_w$ to balance these counteracting effects but this is outside the scope of this work. Instead, $\Delta T_w$ was chosen at the relatively low value of 35 K to reduce energy losses from COP changes and heat transfer so that they can be neglected. With the hot water storage assumed to be inside the living area, heat dissipation may be considered to contribute to space heating anyway. The electricity price was assumed to be 50\$/MWh to be consistent with \citet{schmidt2019projecting}.

\begin{table}[htbp]
    \centering
    \resizebox{0.5\linewidth}{!}{%
    \begin{tabular}{l l l l l l}
        \hhline{======}
         Scheme & Parameter & Symbol & Unit & Value & Source \\
         \hline
         V2G \& smart & (Dis-)charging efficiency & $\eta$ & \% & 92 & \citep{zhang2016evaluation}\\
         charging & Base plug-in time (avg) & $t^{plug-in}_{base}$ & h & 11.5 & \citep{dftEVdata}\\
         & EV battery's energy capacity & $E_{EV}^{max}$ & kWh & 60 & \citep{dftBATdata}\\ 
         & Charger's power capacity & $P_{cha}^{cap}$ & kW & 7.4 & \citep{ukevse}\\
         & Guaranteed minimum charge & $a_{GMC} $ & \% & 30 & \citep{octopus_powerpack} \\
         & Daily driving energy & $E^{daily}_{drive}$ & kWh/day & 5.56 & \citep{dft_mileage,kwhpermile}\\
         & RPT for $p^{WTA}_{base}$ & $\Delta t^{plug-in}_{base}$ & h &  11.5 & \citep{dftEVdata,octopus_powerpack}\\
         & \makecell[tl]{Fraction of energy\\charged at home} & $a_{HC}$ & \% & 90 & \citep{dft_homecharge} \\
         Heat pump & Average power consumption & $P_{HP}^{avg}$ & kW & 0.46 & \citep{lowe2017}\\
         & Average power when on & $P^{avg,act}_{HP}$ & kW & 1.68 & \citep{lowe2017}\\
         & \makecell[tl]{Seasonal performance\\factor} & $SPF$ &  & 2.71  & \citep{lowe2017}\\ 
         & Building's heat capacity & $m^bc_p^b$ & kJ/K & 34,780 & \citep{housing2024,panao2019measured}\\
         Storage & Insulation thickness & $L$ & m & 0.05 & \citep{ARMSTRONG2014128}\\
         & Min. ceiling height & $H$ & m & 2.3 & \citep{dclg_ceilingheight} \\
         & Water density & $\delta^w$ & kg/m\textsuperscript{3} & 1,000 & \citep{luerssen2021global}\\
         & \makecell[tl]{Specific heat\\capacity of water} & $c_p^w$ & kJ/(kg $\cdot$ K) & 4.18 & \citep{luerssen2021global}\\
         & \makecell[tl]{Water storage\\temperature difference} & $\Delta T^w_{max}$ & K & 35 & \makecell[tl]{Design\\choice}\\
         \hhline{======}
    \end{tabular}}
    \caption{Additional Parameters}
    \label{tab:tech_params}
\end{table}

The value factor for the EV-based tariffs is estimated based on data from the Department for Transport \citep{dftEVdata}. For the heat pump-based tariffs, the estimations are based on data from the Renewable Heat Premium Payment scheme \citep{lowe2017}. Price data in both cases comes from the Balancing Mechanism Reporting Service \citep{elecpricedata}.


\section{Results and discussion} \label{sec:res_dis}

\subsection{Value factor}
\begin{figure}
    \centering
    \includegraphics[width=0.5\linewidth]{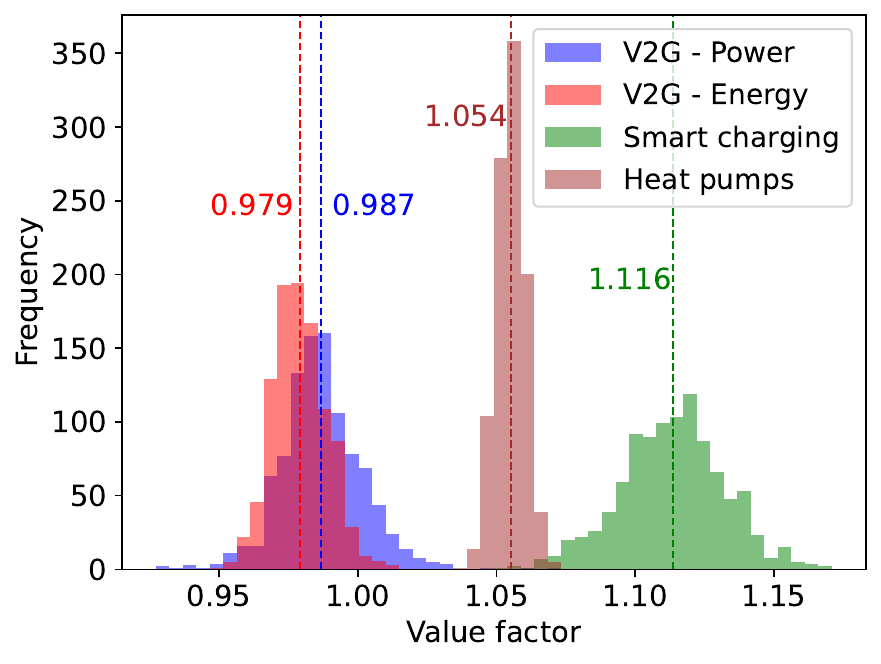}
    \caption{Value factor results with dotted lines indicating averages. The distribution arises from a Monte-Carlo analysis that took sub-samples of 50 DR assets to assess the sensitivity of the value factor to changes in consumer behaviour.}
    \label{fig:vf_res_sens}
\end{figure}

Results for the value factors were obtained using the methodology in Section \ref{Availability adjustments} and the data sources from Section \ref{vfdata}. Figure~\ref{fig:vf_res_sens} shows the value factor results for the different DR schemes. The distribution of value factors in Figure~\ref{fig:vf_res_sens} is linked to variations in the availability profile caused by the subsampling of the Monte-Carlo simulation. It shows that the value factor produces relatively stable estimates even for a low sample size of 50 DR assets. The value factor compares the availability profile of a DR scheme with a continuously available flexibility asset (i.e. most storage technologies). The more a DR asset's availability coincides with high energy prices, the higher its value factor. An asset with an entirely even availability profile by definition has a value factor of one. Value factors above one denote that a DR scheme's availability generally coincides with above-average electricity prices while a value factor below one indicates prevailing availability at times when electricity prices are low. Note that value factors only measure the value of the normalised availability profile. The value factor is unaffected by a DR scheme's total availability hours. These are instead captured by V2G's availability factor ($a_f^{V2G}$) or the average shiftable power consumption of unidirectional assets ($P_{1d}^{avg}$), which, in both cases, determine the required number of assets. For unidirectional DR assets, the availability profile is given by the uncontrolled load profile of the DR asset, as this is the consumption that can be moved by DR. Smart charging has the highest value factor, because uncontrolled domestic charging mostly takes place in the evening hours, coinciding with high demand. Moving domestic EV demand is valuable because it would otherwise occur at times with high energy prices. Heat pump schemes with and without storage have the same value factor because, in both schemes, the uncontrolled demand comes from heat pumps that have the same heating demand profiles in an uncontrolled scenario. Heating demand occurs primarily in the expensive morning and evening hours ($VF>1$) but is generally spread more evenly than smart charging demand, leading to a slightly lower value factor. V2G is the only scheme where the value factor is not determined by the load profiles of uncontrolled demand but by discharge availability instead. Since the used dataset consists mostly of domestic chargers, EVs tend to plug in overnight, covering both the high-demand evening hours as well as low-demand nighttime. This explains their value factor close to unity, meaning their profile value does not vary much from a continuously available asset. Note how there are different V2G-value factors for available energy and power because, differently from other DR schemes, power and energy profiles can diverge for V2G. The power VF is applied when the number of chargers determines the required V2G chargers, while the energy VF is applied when the available battery capacity sets the number of contracted V2G chargers, as determined by Equation (\ref{eq2}).

\begin{figure}[ht]
    \centering
    \includegraphics[width=0.5\linewidth]{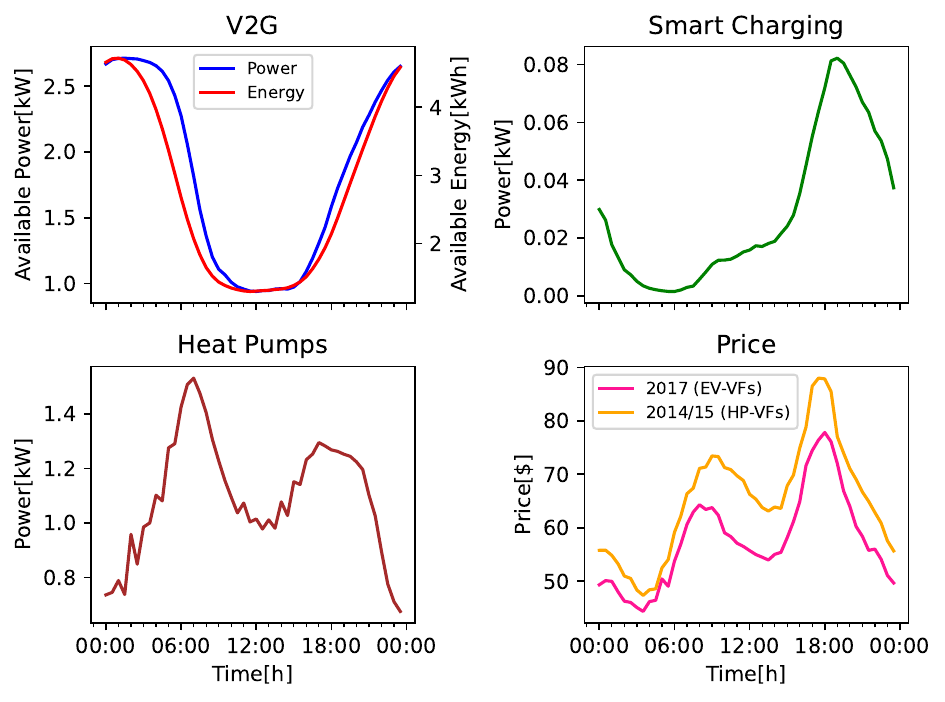}
    \caption{Average daily profiles of value factor parameters}
    \label{fig:vf}
\end{figure}

Figure~\ref{fig:vf} shows daily profiles for the parameters that were used to estimate the different value factors. Comparing the price profile with a parameter's profile gives an idea of how the value factor is estimated. However, it should be noted that seasonal fluctuations in price can naturally also affect the value factor, but cannot be inferred from Figure~\ref{fig:vf}.

\subsection{Levelised cost}

\begin{figure}[ht]
    \centering
    \includegraphics[width=0.5\linewidth]{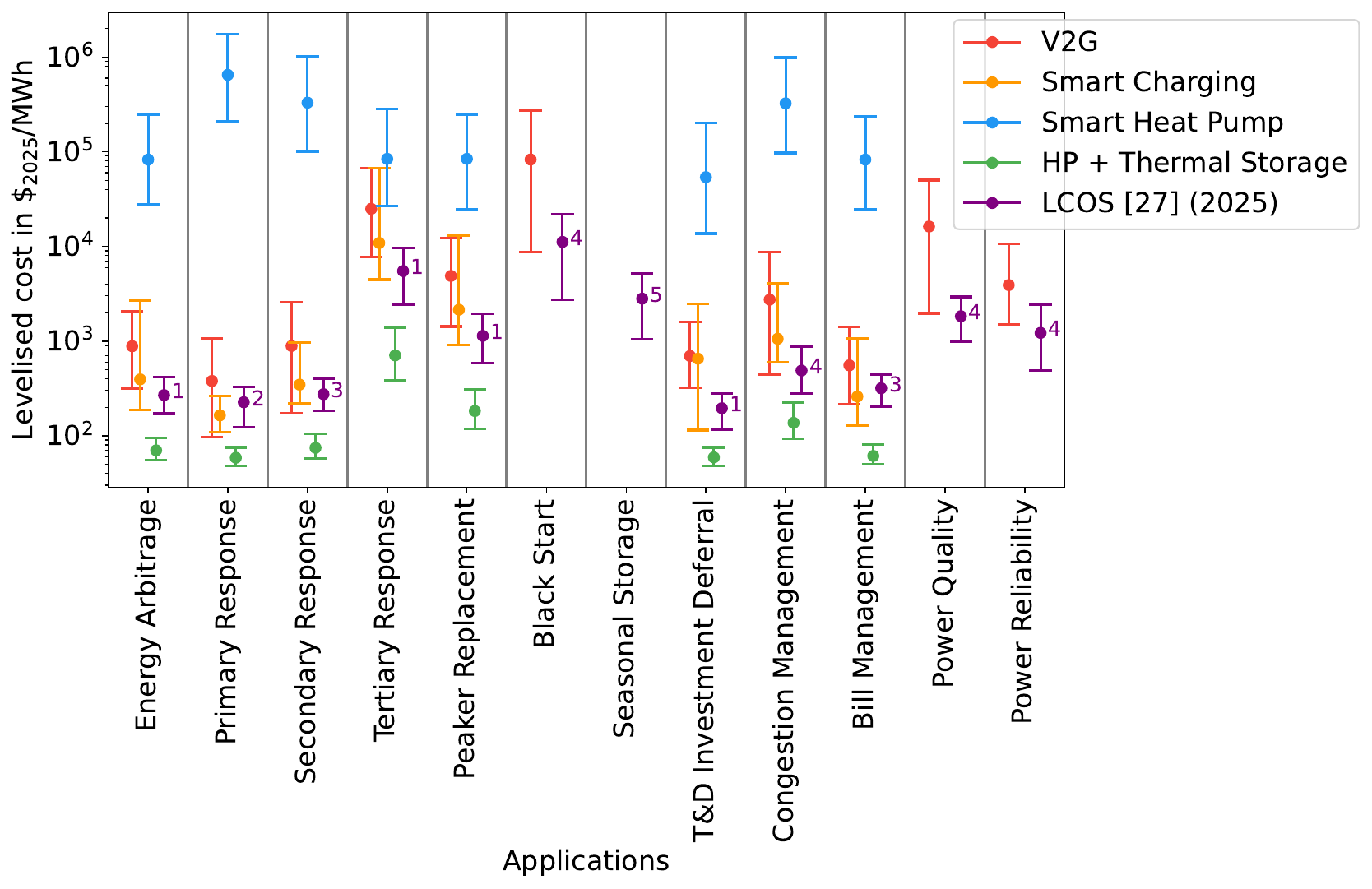}
    \caption{$LCODR_{VF}$ in energy terms from Monte Carlo simulation for the twelve different applications. LCOS results from \citet{schmidt2019projecting} are included for comparison with numbers indicating the cheapest storage technology (1: Pumped hydro, 2: Flywheel, 3: Vanadium-flow, 4: Lithium-ion, 5: Hydrogen, 6: Compressed air).}
    \label{fig:res_energy_abs}
\end{figure}

\begin{figure}[ht]
    \centering
    \includegraphics[width=0.5\linewidth]{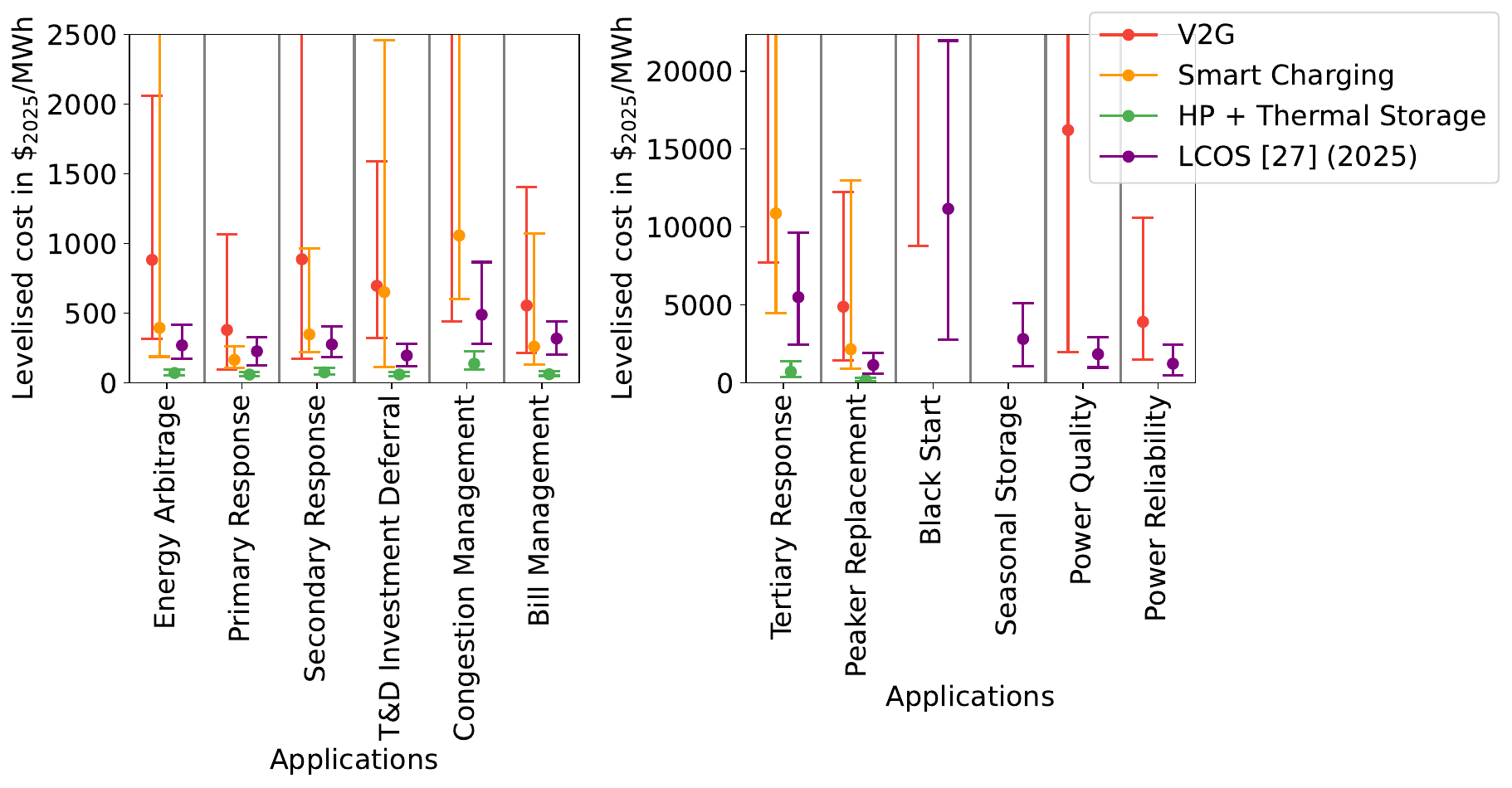}
    \caption{Split-up version of Figure~\ref{fig:res_energy_abs} to allow better results identification on linear scale.}
    \label{fig:res_energy_abs2}
\end{figure}

Figure~\ref{fig:res_energy_abs} and Figure~\ref{fig:res_energy_abs2} show the LCODR in energy terms adjusted for their respective value factor ($LCODR_{VF}$) for the twelve different applications. It also contrasts the $LCODR_{VF}$ results with the cost-optimal LCOS from \citet{schmidt2019projecting}, which was taken from their predictions for the year 2025. While investment cost developments for most storage technologies predict further cost reductions beyond 2025, the main cost component for DR lies in consumer remuneration, for which future cost developments are difficult to predict. This analysis only shows a snapshot of the year 2025 (DCE data from previous years has been adjusted for inflation). When interpreting the cost gradient between DR and storage, it is important to remember that storage technologies are forecast to further reduce their capital costs with time, as cumulative capacity increases past 2025. \citet{schmidt2019projecting} predicts the LCOS of Lithium-ion batteries to reduce by another 40\% until 2050, for example. Additionally, emerging storage technologies, such as Sodium-ion batteries, may further fuel this trend \citep{vaalma2018cost}. The costs of DR, on the other hand, are not predicted to change significantly, and rising incomes may even increase them. While this means that DR's cost-competitiveness is poised to decrease, increased DR penetration in the near future may slow storage capacity installations, in turn reducing the speed of cost reductions for energy storage. DR technologies are included in Figure~\ref{fig:res_energy_abs} following their technical suitability, which is indicated in Table~\ref{tab:my_label}. The results show that heat pumps combined with thermal storage, where feasible, significantly outperform other DR or storage technologies. This underlines a huge potential for cheap flexibility from residential thermal storage. While outside the scope of this study, it should be noted that district heating with large-scale thermal storage may present an even more cost-effective flexibility source, and \citet{vilen2024seasonal} suggests that these could even provide seasonal load-shifting. The smart heat pump scheme cannot compete with either storage or any of the other DR technologies. The EV-based schemes are able to contend with the most cost-competitive storage technology for applications with a high total annual discharge ($N_{app}^{cyc} \times \Delta t_{DD}$), most prominently energy arbitrage, primary response, transmission and distribution (T\&D) investment deferral, and bill management. The reason for this is that higher discharge requirements allow the EV-based schemes to leverage their advantage over storage technologies with lower roundtrip efficiencies.

\begin{figure}[ht]
    \centering
    \includegraphics[width=0.5\linewidth]{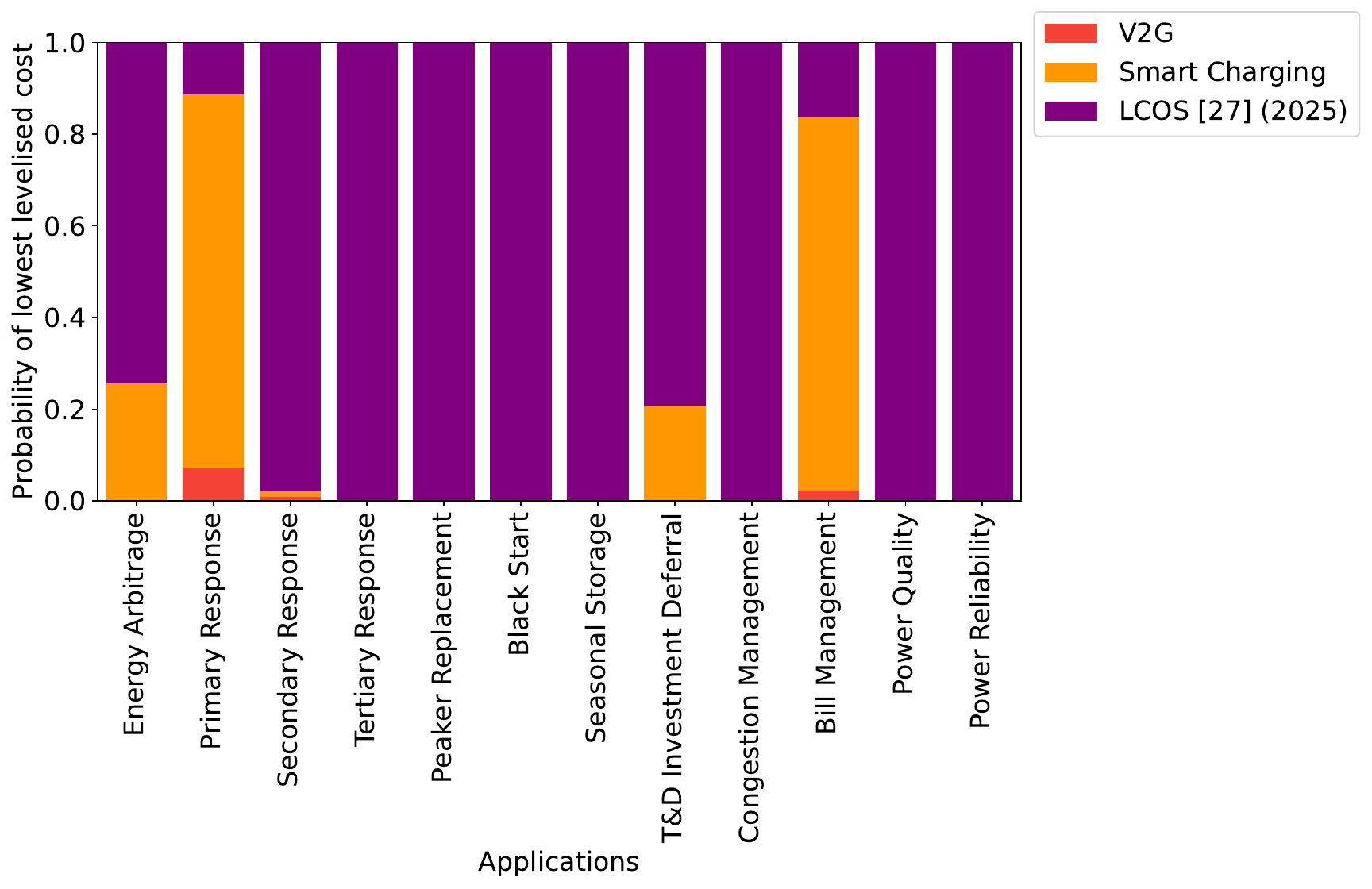}
    \caption{The probabilities of different technologies to achieve the lowest $LCODR_{VF}$ in energy terms for all twelve applications. Results for heat pumps with thermal storage were excluded because these present the cheapest option with a 100\% probability for all applications where they are feasible.}
    \label{fig:res_energy_probs}
\end{figure}

Figure~\ref{fig:res_energy_probs} shows the estimated likelihood of different technologies presenting the cheapest option in terms of energy-based LCODR. It excludes the option of heat pumps with thermal storage, as these would otherwise present the cheapest technology across all applications where they can be used. The equivalent results for the power-based LCODR can be found in the supplementary material (S8). Figures~\ref{fig:res_energy_abs} and \ref{fig:res_energy_probs} show that DLC schemes with the average customer are often unable to provide more cost-effective flexibility than storage assets. Heat pumps with thermal storage are the only one out of the four trialled schemes that achieve consistently lower levelised costs than storage technologies. This insight should encourage energy system stakeholders to pursue flexibility from thermal storage in buildings with electrical heating. EV-based DR schemes cannot compete with thermal storage, but they may still be cost-competitive in applications with a large number of annual activation hours. The deployment of EV-based DR schemes should therefore focus on those applications as they are more likely to outcompete storage this way.

\begin{figure}[t]
    \centering
    \includegraphics[width=0.5\linewidth]{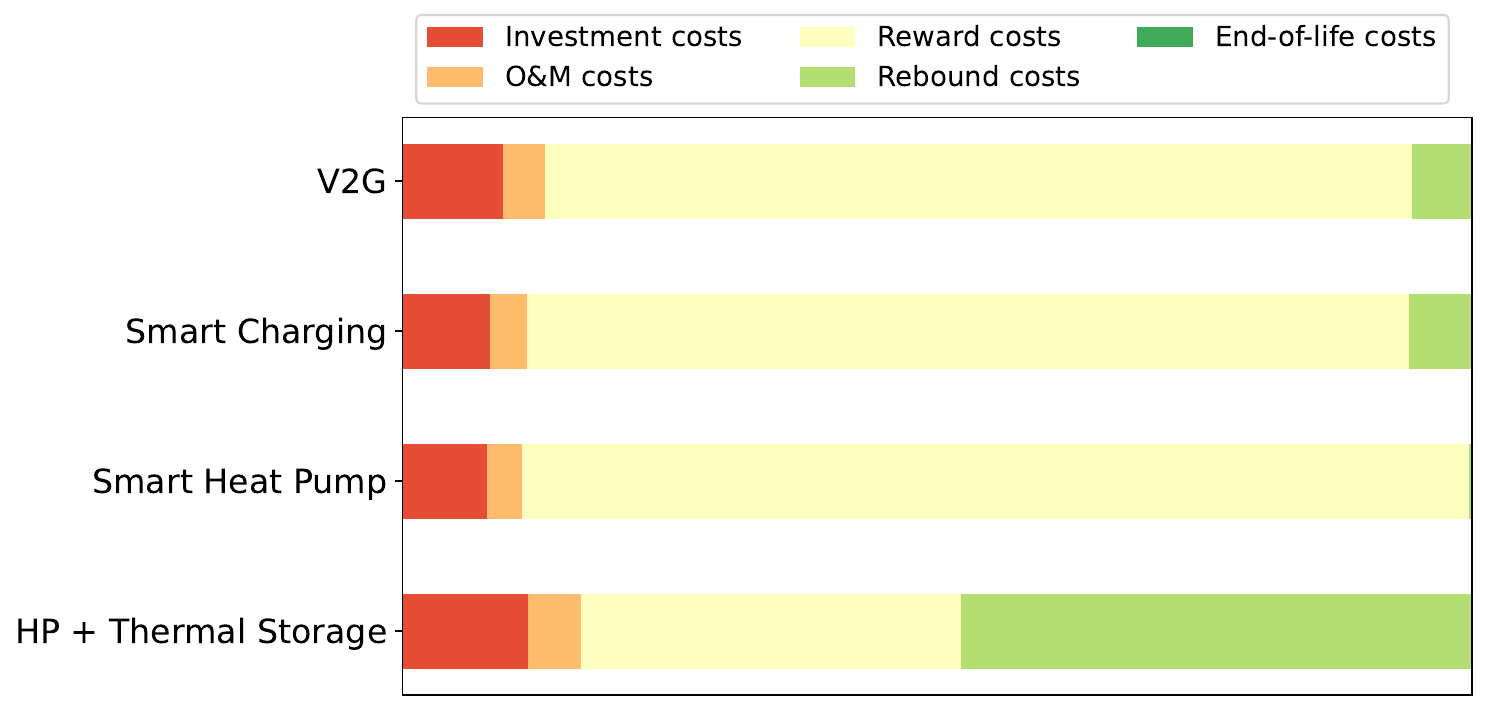}
    \caption{Breakdown of average cost components for different technologies. Note that the share of rebound and reward costs can vary significantly depending on the application, and this figure only shows an average of all applications.}
    \label{fig:cost_compos}
\end{figure}

Figure~\ref{fig:cost_compos} shows the fraction of the LCODR that is associated with each term in the numerator of Equation (\ref{eqb1}). It shows that reward payments make up a large part of the total cost for all studied schemes, highlighting the importance of adequately portraying this cost component in cost assessments of DR. Rebound costs make up a larger part of the thermal storage scheme because of its comparatively low reward payments term.

\subsection{Limitations}
Three major limitations should be considered when interpreting the results.

\begin{enumerate}[leftmargin=*]
    \item DR is not storage. The LCODR methodology builds upon the LCOS methodology because DR can provide the same service as storage in many situations. For storage, however, the times of charge and discharge are more independent of each other than for most of the DR applications, where the rebound sometimes has to occur immediately after the end of the discharge period. The proposed value factor does not account for time constraints on the rebound period, and a comprehensive value assessment should also include a value factor for the energy that is used for the rebound. The heat pump schemes, for example, may have a higher value factor because their consumption is higher in winter, but without the ability to move their consumption seasonally, this should arguably not increase their assigned value. Furthermore, the considered DR resources involve a multitude of distributed assets, while energy storage can be built in large centralised projects, which may affect costs. This aspect has been shown to create a difference in costs between rooftop and utility-scale solar \citep{tsuchida2015comparative}. The distributed nature of DR further suggests that reliability is going to be lower than that of energy storage, even when DR assets are theoretically available, because communication faults are more likely.
    \item V2G also includes a period of Smart Charging. Figure~\ref{fig:ex} shows that V2G vehicles are charged immediately to then discharge as required. However, this period of immediate charging could also be shifted like in the Smart Charging schemes, thereby potentially offering further services from V2G and lowering its costs.
    \item Better data is needed. Several assumptions had to be made due to lacking data, and existing data had to be interpreted outside of its geographic and technical context. To get a measure of how RPT affects discharge duration, for example, it had to be assumed that vehicles plug in once a day and that consumers never plug in for longer than they are contractually obliged to. Rather than measuring the proportion of the day that drivers are plugged in for, future studies should aim to measure the plug-in durations of drivers who have been incentivised financially to plug in their EVs. More granular data may also allow the estimations to move beyond averages and take into account the spectrum of participants in each DR scheme. The consumer data is, furthermore, based on stated-preference surveys, which may suffer from hypothetical bias, as discussed in the supplementary material (S7). Collecting revealed preference data would enable a more accurate and unbiased estimation of the LCODR.    
\end{enumerate}

 \section{Conclusion} \label{sec:concl}
Assessment of lifetime costs of demand response has received relatively little attention, and so this paper addresses the issue by introducing the levelised cost of demand response, LCODR. A methodology for estimating the lifetime costs of storage-equivalent DR services is elaborated across twelve applications that may be provided by storage or DR. Data from academic literature and industry is applied to assess the LCODR for four different demand response schemes. The results suggest that residential heat pumps with thermal storage can provide flexibility at a lower cost than previously studied storage technologies for all applications that can be serviced through demand reduction. Smart charging and, to a lesser extent, vehicle-to-grid can compete with storage technologies in applications that are activated for a comparatively larger proportion of the year, e.g. primary response. Results should be interpreted with caution because data was collated from various sources, with simplifying assumptions required at times. Future work should therefore aim to collect consistent consumer data that is tailored towards a granular LCODR estimation. Additionally, emerging load control schemes for EVs and heat pumps should be targeted as a source for revealed preference data that can be collected at a higher frequency. Extensions of the proposed framework could also consider DR schemes where consumers play a more active role, like household appliances or industrial loads. The proposed methods and results present a versatile decision-making framework for energy system stakeholders who want to compare the costs of DR and storage. To price in the potential environmental advantages of DR over energy storage, future work should extend the comparison framework to consider social costs.

\bibliographystyle{elsarticle-num-names} 
\bibliography{cas-refs}





\end{document}